\documentclass{ws-procs975x65}

\usepackage{slashed}
\usepackage{verbatim}

\begin{document}

\begin{flushright}
MPP-2010-50
\end{flushright}
\vspace{-0.1cm}

\title{NEUTRINO DECAY INTO FERMIONIC QUASIPARTICLES IN LEPTOGENESIS}

\author{CLEMENS P.~KIE\ss IG$^\star$, MICHAEL PL\"UMACHER $^\heartsuit$}

\address{
Max-Planck-Institut f\"ur Physik (Werner-Heisenberg-Institut), F\"ohringer Ring 6\\
D-80805 M\"unchen, Germany\\
$^\star$E-mail: ckiessig@mpp.mpg.de,\\
$^\heartsuit$E-mail: pluemi@mpp.mpg.de
}

\author{MARKUS H.~THOMA$^\dagger$}

\address{Max-Planck-Institut f\"ur extraterrestrische Physik, 
Giessenbachstra\ss e,\\
D-85748 Garching, Germany\\
$^\dagger$E-mail: mthoma@mpe.mpg.de
}

\begin{abstract}
We calculate the decay rate of the lightest heavy Majorana neutrino in
a thermal bath using finite temperature cutting rules and effective
Green's functions according to the hard thermal loop resummation
technique. Compared to the usual approach where thermal masses are
inserted into the kinematics of final states, we find that deviations
arise through two different leptonic dispersion relations. The decay
rate differs from the usual approach by more than one order of
magnitude in the temperature range which is interesting for the weak
washout regime. This work summarizes the results of
Ref.~\refcite{Kiessig:2010pr}, to which we refer the interested
reader.
\end{abstract}


%\classification{11.10.Wx, 13.35.Hb, 14.60.St, 98.80.Cq}

\bodymatter

\section{Introduction}

Leptogenesis~\cite{Fukugita:1986hr} is an extremely
successful theory in explaining the baryon asymmetry of the universe
by adding three heavy right-handed neutrinos $N_i$ to the standard
model,
\begin{equation}
\delta {\mathcal L} = i \bar{N}_i \partial_\mu \gamma^\mu N_i - 
\lambda_{\nu,i
  \alpha} \bar{N}_i \phi^\dagger \ell_\alpha - \frac{1}{2} M_i
\bar{N}_i N_i^c + h.c. \: ,
\end{equation}
with masses $M_i$ at the scale of grand unified theories (GUTs) and
Yukawa couplings $\lambda_{\nu,i \alpha}$ similar to the other
fermions. This also solves the problem of the light neutrino masses
via the see-saw mechanism without fine-tuning~\cite{Minkowski}.

The heavy neutrinos decay into lepton and Higgs boson after inflation,
the decay is out of equilibrium since there are no gauge couplings to
the standard model. If the CP asymmetry in the Yukawa couplings is
large enough, a lepton asymmetry is created by the decays which is
then partially converted into a baryon asymmetry by sphaleron
processes. As temperatures are high, interaction rates and the CP
asymmetry need to be calculated using thermal field
theory~\cite{Giudice:2003jh} rather than vacuum quantum field theory.

\section{Hard Thermal Loops and Thermal Masses}

When using bare thermal propagators in TFT~\cite{LeBellac:1996}, one
can encounter IR singularities and gauge dependent results. In order
to cure this problem, the hard thermal loop (HTL) resummation
technique has been invented~\cite{Braaten:1989mz,Braaten:1990az}. If
$g$ is the coupling to the thermal bath, then for soft momenta $K
\lesssim g T$, resummed propagators are used. For a scalar field
e.g.~this reads
\begin{equation}
i \Delta^*=i \Delta+i \Delta (-i \Pi) i \Delta + \dots =
\frac{i}{\Delta^{-1} -\Pi} = \frac{i}{Q^2-m_0^2-\Pi}.
\end{equation}
The self energy $\Pi \sim g T$ then acts as a thermal mass  $m_{\rm th}^2=\Pi$ 
and gives a correction 
$m_{\rm tot}^2:= m_0^2+m_{\rm th}^2$.

\section{Decay and Inverse Decay Rate}
\label{decay}

Since we are interested in regimes where both the Higgs boson and the
lepton momentum can be soft, we resum both propagators.
The HTL resummation technique has been considered in
Ref.~\refcite{Thoma:1994yw} for the case of a Dirac fermion with
Yukawa coupling.

In order to calculate the interaction rate $\Gamma$ of $N
\leftrightarrow \ell \phi$, we cut the $N$ self energy and use the HTL
resummation for the lepton and Higgs boson propagators. Since
$\lambda_{\nu,i \alpha} \ll 1$, it is justified to neglect the
coupling of the neutrino to the thermal bath.
According to finite-temperature cutting
rules~\cite{Weldon:1983jn,Kobes:1986za}, the interaction rate reads
\begin{equation}
\Gamma(P) = - \frac{1}{2 p_0} \; {\rm tr} [ (\slashed{P}+M) \; {\rm Im} \; 
\Sigma(P)].
\end{equation}
At finite temperature, the self-energy reads
\begin{equation}
\Sigma(P)=-g^2 T \sum_{k_0=i (2 n+1) \pi T} \int \frac{{\rm d}^3 k}{(2
\pi)^3} \: P_L \: S^*(K) \: P_R \: D^*(Q),
\end{equation} 
where $P_L$ and $P_R$ are the projection operators on left- and
right-handed states, $Q=P-K$ and we have summed over neutrino and
lepton spins.  We also sum over the two components of the doublets,
particles and antiparticles and the three lepton flavors, such that
$g^2=4 (\lambda_\nu^\dagger \lambda_\nu)_{11}$.

The HTL-resummed Higgs boson propagator is
\mbox{$D^*(Q)=1/(Q^2-m_\phi^2)$}, where {$m_\phi^2/T^2=(3/16 \, g_2^2+
  1/16 \, g_Y^2 + 1/4 \, y_t^2 + 1/2 \, \lambda)$} is the thermal mass
of the Higgs boson. The couplings denote the SU(2) coupling $g_2$, the U(1)
coupling $g_Y$, the top Yukawa coupling $y_t$ and the Higgs boson self
coupling $\lambda$, where we assume a Higgs boson mass of \mbox{$115$ GeV}. The
other Yukawa couplings can be neglected since they are much smaller
than unity and the remaining couplings are renormalized at the first
Matsubara mode $2 \pi T$ as explained in Ref.~\refcite{Giudice:2003jh}. 

The effective lepton propagator in the helicity-eigenstate
representation is given by~\cite{Braaten:1990wp}
\begin{equation}
\label{fermprop}
S^*(K)=\frac{1}{2} \Delta_+(K) (\gamma_0-\hat{\bf k} \cdot
\boldsymbol{\gamma}) +\frac{1}{2} \Delta_-(K) (\gamma_0+\hat{\bf k} \cdot
\boldsymbol{\gamma}),
\end{equation}
where 
\begin{equation}
\Delta_\pm(K)=\left [ -k_0 \pm  k + \frac{m_\ell^2}{k} \left ( \pm1 - 
\frac{\pm k_0 - k}{2k} \ln \frac{k_0+k}{k_0-k}  \right ) \right ]^{-1}
\end{equation}
and
$m_\ell^2/T^2=(3/32 \, g_2^2+ 1/32 \, g_Y^2)$.

The trace can be evaluated as
\begin{equation}
{\rm tr} [(\slashed{P} +M) P_L S^*(K) P_R]= \Delta_+ (p_0-p \eta)+\Delta_-
(p_0 +p \eta),
\end{equation}
where $\eta={\bf p \cdot k}/pk$ is the angle between neutrino and
lepton. We evaluate the sum over Matsubara frequencies by using the
Saclay method~\cite{Pisarski:1987wc}. For the Higgs boson propagator,
the Saclay representation reads
\begin{equation}
D^*(Q)=-\int_0^\beta {\rm d} \tau e^{q_0 \tau} \frac{1}{2 \omega_q}
\{ [ 1+n_B(\omega_q)] e^{-\omega_q \tau} + n_B(\omega_q) e^{\omega_q \tau}\},
\end{equation}
where \mbox{$\beta=1/T$}, \mbox{$n_B(\omega_q)=1/(e^{\omega_q \beta}-1)$}
is the Bose-Einstein distribution and \mbox{$\omega_q^2=q^2+m_\phi^2$}. For
the lepton propagator it is convenient to use the spectral
representation~\cite{Pisarski:1989cs} 
\begin{equation}
\Delta_\pm(K)=-\int_0^\beta {\rm d \tau'} e^{k_0 \tau'} 
\int_{-\infty}^{\infty}
{\rm d} \omega \: \rho_\pm(\omega,k) [1-n_F(\omega)] e^{-\omega
\tau'},
\end{equation} 
where \mbox{$n_F(\omega)=1/(e^{\omega \beta}+1)$} is the Fermi-Dirac
distribution and $\rho_\pm$ the spectral density~\cite{Braaten:1990wp}.

The lepton propagator in Eq.~\eqref{fermprop} has two different poles
for $1/\Delta_\pm=0$, which correspond to two leptonic quasiparticles
with a positive ($\Delta_+$) or negative ($\Delta_-$) ratio of
helicity over
chirality~\cite{Klimov:1981ka,Weldon:1982bn,Weldon:1989ys}. The
spectral density $\rho_\pm$ has a contribution from the poles and a
discontinuous part. We are interested in the pole contribution
\begin{equation}
\label{rho}
\rho_\pm^{\rm pole}(\omega,k)=\frac{\omega^2-k^2}{2 m_\ell^2} (\delta(\omega-
\omega_\pm)+ \delta (\omega+ \omega_\mp)),
\end{equation}
where $\omega_\pm$ are the dispersion relations for the two
quasiparticles, i.e.~the solutions for $1/\Delta\pm(\omega_\pm,{\bf
k})=0$, shown in Fig.~\ref{modmass}~(a). An analytical solution for
$\omega_\pm$ can be found in the appendix of
Ref.~\refcite{Kiessig:2010pr}.
One can assign a momentum-dependent thermal mass
\mbox{$m_\pm(k)^2=\omega_\pm(k)^2-k^2$} to the two modes as shown in
Fig.~\ref{modmass}~(b) and for very large momenta the heavy mode $m_+$
approaches $\sqrt{2} \: m_\ell$, while the light mode becomes
massless.  
\def\figsubcap#1{\par\noindent\centering\footnotesize(#1)}
\begin{figure}[b]%
\label{modes}
\begin{center}
\hspace{-0.25cm}
 \parbox{6.05cm}{\label{modes}\epsfig{figure=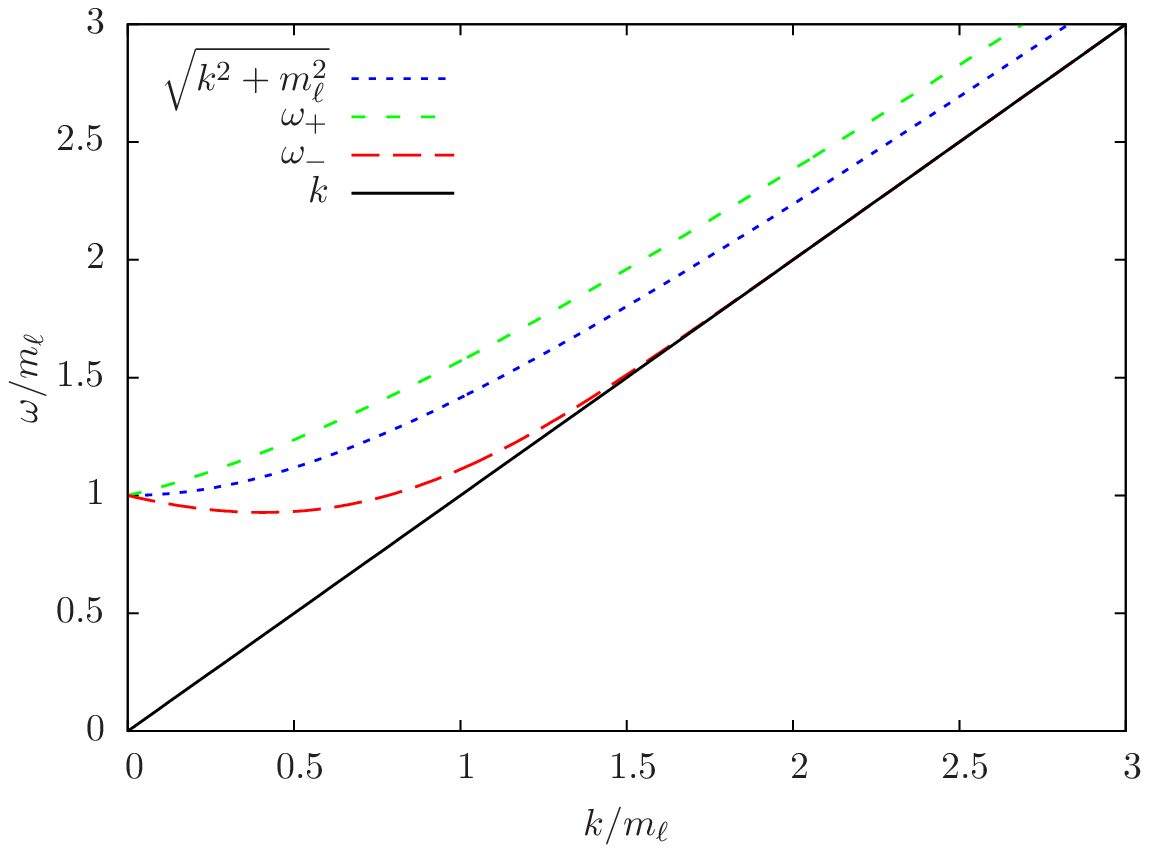,width=6.25cm}
 \figsubcap{a}\label{modes}}
 \hspace*{0.2cm}
 \parbox{6.05cm}{\epsfig{figure=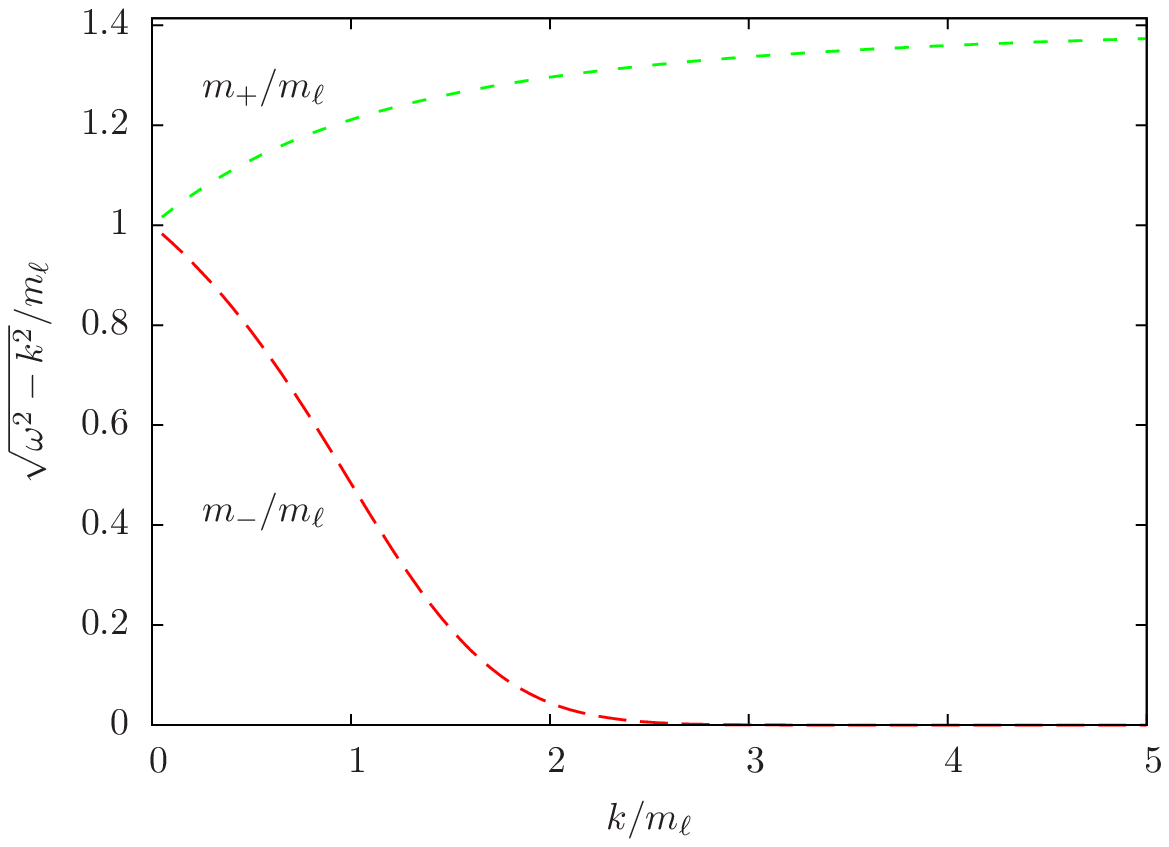,width=6.25cm}
 \figsubcap{b}\label{masses}}
 \caption{(a) The two leptonic dispersion relations compared with the
   standard dispersion relation. (b)
   The momentum-dependent quasiparticle masses
   $m_\pm^2=\omega_\pm^2-k^2$.}
\label{modmass}
\end{center}
\end{figure}

After evaluating the sum over $k_0$, carrying out the integrations
over $\tau$ and $\tau'$ and integrating over the pole part of
$\rho_\pm$ in Eq.~\eqref{rho}, we get
\begin{equation}
\begin{split}
T \sum_{k_0} D^* \Delta_\pm=-\frac{1}{2 \omega_q} 
& \left \{ 
\frac{\omega_\pm^2-k^2}
{2 m_\ell^2} \left [ 
\frac{1+n_B-n_F}{p_0-\omega_\pm-\omega_q}+
 \frac{n_B+n_F}{p_0-\omega_\pm+\omega_q} 
\right ]
\right. \\
& 
\label{frequencysum}
\left. +  \frac{\omega_\mp^2-k^2}{2 m_\ell^2} \left [ 
\frac{n_B+n_F}{p_0+\omega_\mp-\omega_q}+
 \frac{1+n_B-n_F}{p_0+\omega_\mp+\omega_q} \right ] 
\right \},
\end{split}
\end{equation} 
where $n_B=n_B(\omega_q)$ and $n_F=n_F(\omega_\pm)$ or
$n_F(\omega_\mp)$, respectively.

The four terms in Eq.~\eqref{frequencysum} correspond to the processes
with the energy relations indicated in the denominator, i.e.~the decay
$N \rightarrow \phi \ell$, the production $N \phi \rightarrow \ell$,
the production $N \ell \rightarrow \phi$ and the production of $N \ell
\phi$ from the vacuum, as well as the four inverse 
reactions~\cite{Weldon:1983jn}. We are only interested in the process
$N \leftrightarrow \phi \ell$, where the decay and inverse decay are
illustrated by the statistical factors $1+n_B-n_F=(1+n_B)(1-n_F)+n_B
n_F$, given by the first term of Eq.~\eqref{frequencysum}.

For carrying out the integration over the angle $\eta$, we use
\begin{equation}
{\rm Im} \frac{1}{p_0-\omega_\pm-\omega_q}= - \pi
\delta(p_0-\omega_\pm-\omega_q) = - \pi \frac{\omega_q}{k p}
\delta(\eta-\eta_\pm). 
\end{equation}
After integrating over $\eta$ we get
\begin{equation}
\begin{split}
\Gamma(P)
=& \frac{1}{16 \pi p_0 p} \sum_\pm \int_{-1 \leq \eta_\pm \leq 1} {\rm d} k 
\, \frac{k}{\omega_\pm} \; |\mathcal{M}_\pm(P,K)|^2 
[1+n_B(\omega_{q \pm}) -n_F(\omega_\pm)] \\ 
=& \frac{1}{2 p_0}\int \, {\rm d} \tilde{k} \, {\rm d} \tilde{q} \; 
(2 \pi)^4 \delta^4(P-K-Q) \; |\mathcal{M}_\pm(P,K)|^2 
 [1+n_B-n_F],
\end{split}
\end{equation}
where $\omega_{q \pm}=p_0-\omega_\pm$,  we only integrate over
regions with $-1 \leq \eta \leq 1$,
${\rm d} \tilde{k}={\rm d}^3 k/((2 \pi)^3 2 \,k_0)$
and ${\rm d}\tilde{q}$ analogously and the matrix elements are
\begin{equation}
|\mathcal{M}_\pm(P,K)|^2=g^2 \frac{\omega_\pm^2-k^2}{2 m_\ell^2} \omega_\pm 
\left (p_0 \mp p \eta_\pm \right ). 
\end{equation}

In order to compare our result to the conventional
approximation~\cite{Giudice:2003jh}, we do the same calculation for an
approximated lepton propagator \mbox{$S^*_{\rm
  approx}(K)=1/(\slashed{K}-m_\ell)$}. This amounts to setting
\mbox{$\omega^2=k^2+m_\ell^2$}, \mbox{$\omega_q=p_0-\omega$} and we get
$|\mathcal{M}|^2=\frac{g^2}{2} (M^2+m_\ell^2-m_\phi^2)$ as matrix
element.

This result resembles the zero temperature result
with zero temperature masses $m_\ell$, $m_\phi$.
The missing factor
$1+n_B-n_F=(1+n_B)(1-n_F)+n_B n_F$
accounts for the statistical distribution of the initial or final
particles.  As pointed out in more detail in Ref.~\refcite{Kiessig:2009cm}, we
have shown that the approach to treat thermal masses like zero
temperature masses in the final state~\cite{Giudice:2003jh} is
justified since it equals the HTL treatment with an approximate
lepton propagator. However this approach does not equal the full HTL
result.

\section{Decay Density}
\label{leptogenesis}

The quantity which enters the Boltzmann equations is the decay density
integrated over all neutrino momenta. In equilibrium it reads
\begin{equation}
\gamma_D^{\rm eq}= \int \frac{{\rm d^3} p}{(2 \pi)^3} f_N^{\rm eq}(E) \, 
\Gamma_D= \frac{1}{2 \pi^2} \int_M^\infty {\rm d} E \: E \, p \, f_N^{\rm eq} 
\, 
\Gamma_D,
\end{equation} 
where $E=p_0$, $f_N^{\rm eq}(E)=[\exp(E \beta)-1]^{-1}$ is the
equilibrium distribution of the neutrinos and $\Gamma_D=[1-f_N^{\rm
eq}(E)] \, \Gamma$.  

\begin{figure}
\begin{center}
\includegraphics[height=7.0cm]{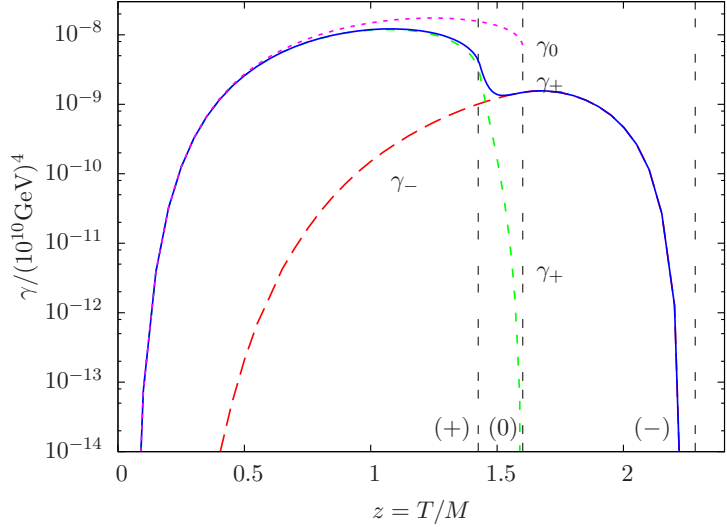}
\caption{\label{comp} The neutrino decay density with the one lepton mode 
approach $\gamma_0$ and the two-mode treatment $\gamma_\pm$ for
$M_{1}=10^{10}$ GeV and $\tilde{m}_1 = 0.06$ eV. The thresholds for
the two modes (+), (-) and one mode (0) are indicated.}
\end{center}
\end{figure}

In Fig.~\ref{comp}, we compare our result to the conventional
approximation~\cite{Giudice:2003jh}. We evaluate the decay rate for
\mbox{$M_1=10^{10}$ GeV} and \mbox{$\tilde{m_1}=(\lambda_\nu
\lambda^\dagger_\nu)_{11} \, v^2/M_1=0.06$ eV}, where \mbox{$v=174$
  GeV} is the vacuum expectation value of the Higgs field.

In the one-mode approach, the decay is forbidden when
$M<m_\ell+m_\phi$. Considering two modes, the phase space is
reduced for the positive mode due to the larger quasi-mass and at
$M=m_+(\infty)+m_\phi$, the decay is only possible into leptons with
small momenta, thus the rate drops dramatically. The decay into the
negative, quasi-massless mode is suppressed due to its much smaller
residue. However, the decay is possible up to $M=m_\phi$. These rates
differ from the one mode approach by more than one order of magnitude
in the interesting temperature regime of $z=T/M \gtrsim 1$.

\section{Conclusions}
\label{conclusions}

As discussed in detail in Ref.~\refcite{Kiessig:2009cm}, we have, by
employing HTL resummation and finite temperature cutting rules,
confirmed that treating thermal masses as kinematic masses as in
Ref.~\refcite{Giudice:2003jh} is a reasonable approximation. We have
calculated the decay density of the lightest heavy Majorana neutrino
and its behavior can be explained by considering the dispersion
relations $\omega_\pm$ of the lepton modes and assigning momentum-dependent
quasi-masses to them. The thresholds for neutrino decay reported in
Ref.~\refcite{Giudice:2003jh} are shifted and the decay density shows
deviations of more than an order of magnitude in the interesting
temperature regime $T/M \sim 1$. In order to arrive at a minimal
consistent treatment, also the decay $\phi \rightarrow N
\ell$ at high temperatures needs to be included as well as Higgs boson 
and neutrino CP asymmetries which are corrected for lepton modes.

This contribution summarizes the results of an earlier
work~\cite{Kiessig:2010pr} and we refer the interested reader to the
more elaborate treatment there.\\[1ex]

{\bf Acknowledgements} We thank Georg
Raffelt, Florian Hahn-W\"ornle, Steve Blanchet, Matthias Garny, Marco
Drewes, Wilfried Buchm\"uller, Martin Spinrath and Philipp Kostka for
fruitful and inspiring discussions.


\begin{thebibliography}{15}

\bibitem{Kiessig:2010pr}
  C.~P.~Kie\ss ig, M.~Pl\"umacher and M.~H.~Thoma,
  %``Decay of a Yukawa fermion at finite temperature and applications to
  %leptogenesis,''
  arXiv:1003.3016 [hep-ph].
  %%CITATION = ARXIV:1003.3016;%%

\bibitem{Fukugita:1986hr} 
  M.~Fukugita and T.~Yanagida, 
%  ``Baryogenesis without grand unification,'' 
  {\it Phys.\ Lett.\ B} {\bf 174}, 45 (1986).
  %%CITATION = PHLTA,B174,45;%%

\bibitem{Minkowski}
  P.~Minkowski, 
%  " Mu $\to$ e gamma at a rate of one out of 1-billion muon decays?,"
  {\it Phys.~Lett.~B} {\bf 67}, 421 (1977).
  %%CITATION = PHLTA,B67,421;%%

\bibitem{Giudice:2003jh}
  G.~F.~Giudice, A.~Notari, M.~Raidal, A.~Riotto and A.~Strumia,
%  ``Towards a complete theory of thermal leptogenesis in the SM and MSSM,'' 
  {\it Nucl.\ Phys.\ B} {\bf 685},89 (2004) [arXiv:hep-ph/0310123].
  %%CITATION = NUPHA,B685,89;%%

\bibitem{LeBellac:1996}
  M.~Le~Bellac, {\it Thermal field theory} (Cambridge University Press,
  Cambridge, UK, 1996).

\bibitem{Braaten:1989mz}
  E.~Braaten and R.~D.~Pisarski,
%  ``Soft amplitudes in hot gauge theories: A general analysis,''
  {\it Nucl.\ Phys.\  B} {\bf 337},569 (1990).
  %%CITATION = NUPHA,B337,569;%%

\bibitem{Braaten:1990az}
  E.~Braaten and R.~D.~Pisarski,
%  ``Deducing hard thermal loops from Ward identities,''
  {\it Nucl.\ Phys.\  B} {\bf 339},310 (1990).
  %%CITATION = NUPHA,B339,310;%%

\bibitem{Thoma:1994yw}
  M.~H.~Thoma,
%  ``Damping of a Yukawa fermion at finite temperature,''
  {\it Z.\ Phys.\  C} {\bf 66}, 491 (1995).
  [arXiv:hep-ph/9406242].
  %%CITATION = ZEPYA,C66,491;%%

\bibitem{Weldon:1983jn}
  H.~A.~Weldon,
%  ``Simple rules for discontinuities in finite temperature field theory,''
  {\it Phys.\ Rev.\  D} {\bf 28}, 2007 (1983).
  %%CITATION = PHRVA,D28,2007;%%

\bibitem{Kobes:1986za}
  R.~L.~Kobes and G.~W.~Semenoff,
%  ``Discontinuities of Green functions in field theory at finite temperature
%  and density. 2,''
  {\it Nucl.\ Phys.\  B} {\bf 272}, 329 (1986).
  %%CITATION = NUPHA,B272,329;%%

\bibitem{Braaten:1990wp}
  E.~Braaten, R.~D.~Pisarski and T.~C.~Yuan,
%  ``Production of soft dileptons in the quark - gluon plasma,''
  {\it Phys.\ Rev.\ Lett.\ }  {\bf 64}, 2242 (1990).
  %%CITATION = PRLTA,64,2242;%%

\bibitem{Pisarski:1987wc}
  R.~D.~Pisarski,
%  ``Computing finite temperature loops with ease,''
  {\it Nucl.\ Phys.\  B} {\bf 309}, 476 (1988).
  %%CITATION = NUPHA,B309,476;%%

\bibitem{Pisarski:1989cs}
  R.~D.~Pisarski,
%  ``Renormalized gauge propagator in hot gauge theories,''
  {\it Physica A} {\bf 158}, 146 (1989).
  %%CITATION = PHYSA,A158,146;%%

\bibitem{Klimov:1981ka}
  V.~V.~Klimov,
%  ``Spectrum of elementary Fermi excitations in quark gluon plasma. (in
%  Russian),''
  {\it Sov.\ J.\ Nucl.\ Phys.\ }  {\bf 33}, 934 (1981)
  [{\it Yad.\ Fiz.\ }  {\bf 33}, 1734 (1981)].
  %%CITATION = YAFIA,33,1734;%%

\bibitem{Weldon:1982bn}
  H.~A.~Weldon,
%  ``Effective fermion masses of order gT in high temperature gauge theories
%  with exact chiral invariance,''
  {\it Phys.\ Rev.\  D} {\bf 26}, 2789 (1982).
  %%CITATION = PHRVA,D26,2789;%%

\bibitem{Weldon:1989ys}
  H.~A.~Weldon,
%  ``Dynamical holes in the quark - gluon plasma,''
  {\it Phys.\ Rev.\  D} {\bf 40}, 2410 (1989).
  %%CITATION = PHRVA,D40,2410;%%

\bibitem{Kiessig:2009cm}
  C.~P.~Kie\ss ig and M.~Pl\"umacher,
%  ``Thermal masses in leptogenesis,''
 [arXiv:hep-ph/0910.4872].
  %%CITATION = ARXIV:0910.4872;%%

\end{thebibliography}
\end{document}